\documentclass[%
 aip, apl,
amsmath,amssymb,
 reprint,%
]{revtex4-1}

\usepackage{graphicx}
\usepackage{dcolumn}
\usepackage{bm}

\usepackage[utf8]{inputenc}
\usepackage[T1]{fontenc}
\usepackage{mathptmx}
\usepackage{color}
\UseRawInputEncoding

\begin{document}

\preprint{AIP/123-QED}

\title[Sample title]{Perfect optical coherence lattices}


\author{Chunhao Liang}
\affiliation{Shandong Provincial Engineering and Technical Center of Light Manipulation \& Shandong Provincial Key Laboratory of Optics and Photonic Devices, School of Physics and Electronics, Shandong Normal University, Jinan 250014, China}
\affiliation{Collaborative Innovation Center of Light Manipulations and Applications, Shandong Normal University, Jinan 250358, China}

\author{Xin Liu}
\affiliation{Shandong Provincial Engineering and Technical Center of Light Manipulation \& Shandong Provincial Key Laboratory of Optics and Photonic Devices, School of Physics and Electronics, Shandong Normal University, Jinan 250014, China}
\affiliation{Collaborative Innovation Center of Light Manipulations and Applications, Shandong Normal University, Jinan 250358, China}

\author{Zhiheng Xu}
\affiliation{Shandong Provincial Engineering and Technical Center of Light Manipulation \& Shandong Provincial Key Laboratory of Optics and Photonic Devices, School of Physics and Electronics, Shandong Normal University, Jinan 250014, China}
\affiliation{Collaborative Innovation Center of Light Manipulations and Applications, Shandong Normal University, Jinan 250358, China}

\author{Fei Wang}
\email{Authors to whom correspondence should be addressed: fwang@suda.edu.cn csuwenwei@126.com yangjiancai@suda.edu.cn pujuanma@foxmail.com}
\affiliation{School of Physical Science and Technology, Soochow University, Suzhou 215006, China} 

\author{Wen Wei}
\email{Authors to whom correspondence should be addressed: fwang@suda.edu.cn csuwenwei@126.com yangjiancai@suda.edu.cn pujuanma@foxmail.com}
\affiliation{College of Mathematics and Physics Science, Hunan University of Arts and Science, Changde 415000, China} 

\author{Sergey A. Ponomarenko}
\affiliation{Department of Electrical and Computer Engineering, Dalhousie University, Halifax, Nova Scotia, B3J 2X4, Canada} 
\affiliation{Department of Physics and Atmospheric Science, Dalhousie University, Halifax, Nova Scotia, B3H 4R2, Canada} 

\author{Yangjian Cai}
\email{Authors to whom correspondence should be addressed: fwang@suda.edu.cn csuwenwei@126.com yangjiancai@suda.edu.cn pujuanma@foxmail.com} 
 \affiliation{Shandong Provincial Engineering and Technical Center of Light Manipulation \& Shandong Provincial Key Laboratory of Optics and Photonic Devices, School of Physics and Electronics, Shandong Normal University, Jinan 250014, China}
\affiliation{Collaborative Innovation Center of Light Manipulations and Applications, Shandong Normal University, Jinan 250358, China}
\affiliation{School of Physical Science and Technology, Soochow University, Suzhou 215006, China} 

\author{Pujuan Ma}
\email{Authors to whom correspondence should be addressed: fwang@suda.edu.cn csuwenwei@126.com yangjiancai@suda.edu.cn pujuanma@foxmail.com} 
\affiliation{Shandong Provincial Engineering and Technical Center of Light Manipulation \& Shandong Provincial Key Laboratory of Optics and Photonic Devices, School of Physics and Electronics, Shandong Normal University, Jinan 250014, China}
\affiliation{Collaborative Innovation Center of Light Manipulations and Applications, Shandong Normal University, Jinan 250358, China}

\date{\today}

\begin{abstract}
We advance and experimentally implement a protocol to generate perfect optical coherence lattices (OCL) that are not modulated by an envelope field. Structuring the amplitude and phase of an input partially coherent beam in a Fourier plane of an imaging system lies at the heart of our protocol. In the proposed approach, the OCL node profile depends solely on the degree of coherence (DOC) of the input beam such that, in principle, any lattice structure can be attained via proper manipulations in the Fourier plane. Moreover, any genuine partially coherent source can serve as an input to our lattice generating imaging system. Our results are anticipated to find applications to optical field engineering and multi-target probing among others.
\end{abstract}

\maketitle
 
Coherent optical lattices, such as periodic structures of optical field amplitude, phase or polarization have long been the focus of attention of the optical community due to a wealth of their applications to the subjects as diverse as neutral gas heating~\cite{1}, coherent manipulation of cold atoms~\cite{2}, cold-atom superfluidity exploration~\cite{3}, and quantum-state control~\cite{4}, to mention but a few examples. Recently, an altogether different kind of optical lattices, optical coherence lattices (OCL), referring to partially coherent light sources with spatially periodic degrees of coherence ~\cite{5}, has triggered much interest ever since their theoretical introduction~\cite{5,6} and subsequent exploration~\cite{7,8,9,10,11,12}. Much interest in OCLs of late has been due to their intriguing propagation characteristics. For instance, the periodicity reciprocity arises between the source coherence and far-zone intensity of OCLs on their free space propagation~\cite{6} which has inspired researchers to realize scalar/vector beam arrays with adjustable spatial distributions and node profiles in the far zone or a focal plane of an imaging system~\cite{7,8,9,11}. The propagation of OCLs through the atmospheric and oceanic turbulence has been studied as well. It was found that the OCLs of higher beam order are less distorted by the turbulence than are the conventional Gaussian Schell-model (GSM) beams~\cite{13,14}. 

Yet, only a few protocols for the experimental realization of OCLs have been reported to date~\cite{15,16}. In Ref.~\cite{15}, an uncorrelated superposition of elementary beams has been utilized to generate OCLs. However, a finite spot size of an elementary beam, led to the appearance of an envelope field $J_1(r)/r$ modulating an OCL produced via this protocol. In Ref.~\cite{16}, OCLs were generated through scattering of light beam by a specially designed random medium which imposed envelope fields on the produced OCLs as well. Hence, the inevitability of the envelope field is germane to all the methods for OCL generation reported in the literature to date. Moreover, although OCLs of different lattice structure can be generated by the protocols of Refs. ~\cite{15,16}, these protocols lack simultaneous control of the lattice node profile and lattice structure. 

In this Letter, we draw on a previously elaborated arsenal of Fourier optics techniques to engineer partially coherent beams~\cite{17,18,19,20,21,22,23,24,25} to advance an efficient protocol to realize OCLs that are not embedded into any envelope field, thereby maintaining a strictly periodic coherence structure that we dub a perfect optical coherence lattice. In our protocol, each OCL node profile depends only on the degree of coherence of a light source, and the lattice structure is determined by the Fourier spectrum of a transmission function of our imaging system. We illustrate our protocol with numerical examples and implement it experimentally. 

 We start by illustrating in Fig. 1 a typical 4f optical system consisting of two identical thin lenses of focal length $f$. We place a plate with a complex transmission function $P(\xi)$ at the rear focal plane of a lens L1. We treat the front focal plane of the lens L1 and the rear focal plane of a lens L2 as the input and output planes to our imaging system, respectively. We generate perfect OCLs in the output plane through modulation of the amplitude and phase of an optical beam in the Fourier plane of the system.

\begin{figure*}[!t]
	\includegraphics[width=12.8cm]{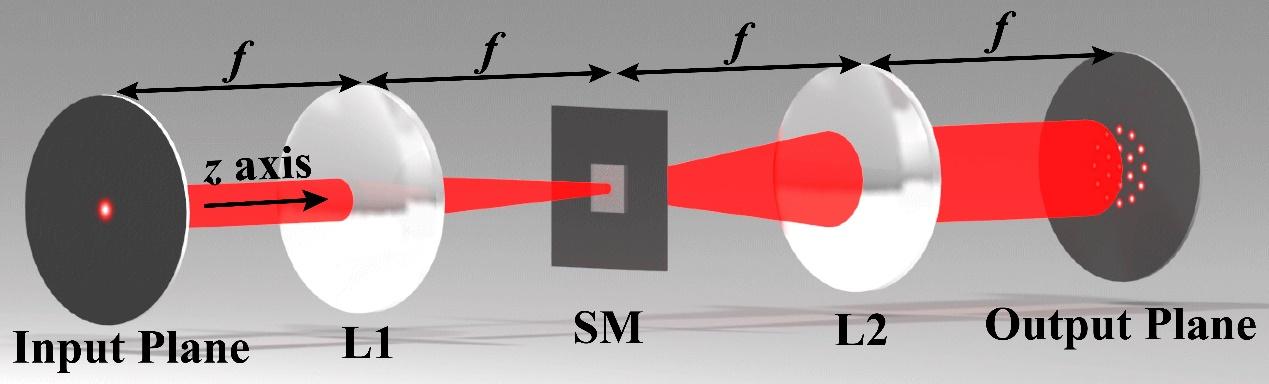}
	\caption{Schematic diagram of an optical system with Fourier phase lattices. L1 and L2, thin lens; SM, spectral modulator.}
	\label{figure1}
\end{figure*}

We assume the input to our system to be a quasi-monochromatic, statistically stationary beam propagating along the $z$ direction. In the space-frequency domain, the second-order statistics of the beam are characterized by the cross-spectral density (CSD) function $W\left( {{\bf{r}_1},{\bf{r}_2},\omega } \right) = \left\langle {{E^*}\left( {{\bold{r_1}},\omega } \right)E\left( {{\bold{r_2}},\omega } \right)} \right\rangle$, where $E$ stands for the random electric field; $\bold{r_1}$ and $\bold{r_2}$ are two position vectors in the input plane; the asterisk and the angle brackets, denote a complex conjugate and ensemble average. Further, $\omega$ is a frequency of light which will be omitted hereafter for brevity. As our imaging system is linear, the input and output CSDs are related through a linear transform as 
\begin{equation}
\begin{aligned}
{W^{\left( {{\rm{out}}} \right)}}\left( {\bold{r}_1^{'},\bold{r}_2^{'}} \right)= \int {{W^{\left( {{\rm{in}}} \right)}}\left( {{\bold{r}_1},{\bold{r}_2}} \right)} {h^*}\left( {{{\bold{r}}_1},\bold{r}_1^{'}} \right)h\left( {{{\bold{r}}_2},\bold{r}_2^{'}} \right){d^2}{\bold{r}_1}{d^2}{\bold{r}_2},
\end{aligned}
\label{eq1}
\end{equation}
where ${W^{\left( {{\rm{in}}} \right)}}$ and ${W^{\left( {{\rm{out}}} \right)}}$ are the input and output CSDs,   $\bold{r}^{'}_i = \left( {x^{'}_i, y^{'}_i} \right),{\rm{ }}i = 1,2$ is an arbitrary position vector in the output plane and  $h\left( {{\bold{r}},{\bold{r}^{'}}} \right)$ is a response function of the optical system. For a 4f optical system shown in Fig. 1, the latter takes the form ~\cite{19} 
\begin{equation}
h\left( {{\bf{r}},{\bf{r}^{'}}} \right) =  - \frac{1}{{{\lambda ^{2}}{f^{2}}}}\int {P\left( \xi  \right)} \exp \left[ { - \frac{{ik}}{f}\xi  \cdot \left( {{\bf{r}} - {\bf{r}^{'}}} \right)} \right]{d^{2}}\xi .
\label{eq2}
\end{equation}
Here $\xi$ is a transverse position vector in the Fourier plane. Eq. (2) indicates that the translationally invariant response function is just a Fourier transform of the complex transmission function $P$. It follows that $h\left( {{\bf{r}},{\bf{r}^{'}}} \right) = h\left( {{\bf{r}} - {\bf{r}^{'}}} \right) = \tilde P\left( {{\bf{r}} - {\bf{r}^{'}}} \right)$, where the tilde denotes the Fourier transform. 

\begin{figure*}
	\includegraphics[width=12.8cm]{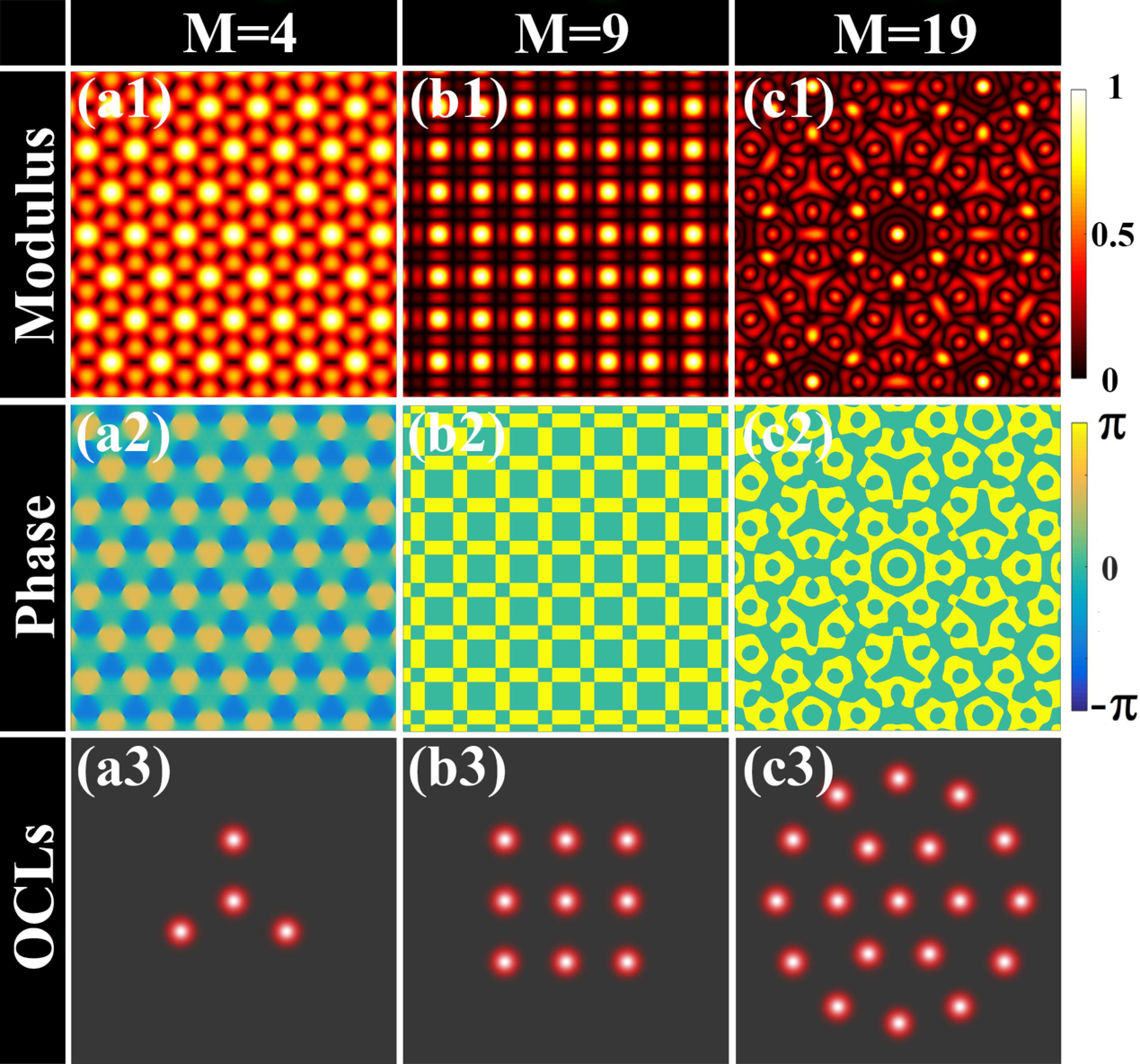}
	\caption{Generation of perfect OCLs (${\left| {{\mu ^{\left( {out} \right)}}\left( {{\bold{r}_1^{'}},{\bold{r}_2^{'}} = 0} \right)} \right|^2}$) of variable lattices structure [(a3)-(c3)] produced via Fourier spectrum modulation. The amplitude and phase distributions of the relevant P functions are displayed in Figs.2 (a1)-(c1) and Figs.2 (a2)-(c2), respectively. The input is a Gaussian Schell-model beam. The distance between the adjacent nodes of each lattice in Fig. 2 (a3)-(c3) is 3 mm.}
	\label{figure2}
\end{figure*}

\begin{figure*}
	\includegraphics[width=12.8cm]{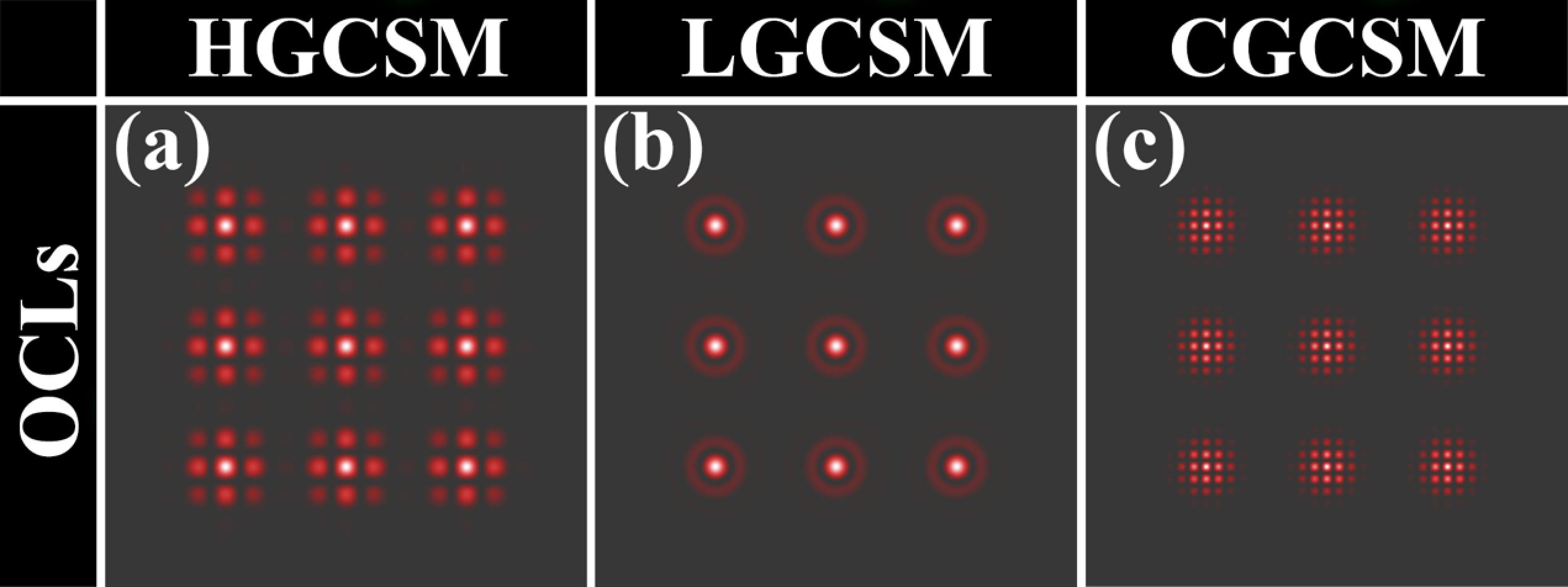}
	\caption{Generation of perfect OCLs (${\left| {{\mu ^{\left( {out} \right)}}\left( {{\bold{r}_1^{'}},{\bold{r}_2^{'}} = 0} \right)} \right|^2}$) of controllable node profiles. Input beams: (a) HGCSM beam with the indices n=m=2, (b) LGCSM beam of the order n=4, and (c) CGCSM beam of the order n=2. The distance between the adjacent nodes of each lattice equals to 3mm.}
	\label{figure3}
\end{figure*}

To realize perfect OCLs in the output plane, we require the response function of the form 
\begin{equation}
h\left( {{\bf{r}} - {\bf{r}^{'}}} \right) =  - \sum\limits_{m = 1}^M {\delta \left( {x - {x^{'}} + ma} \right)\delta \left( {y - {y^{'}} + mb} \right)} ,
\label{eq3}
\end{equation}
where $\delta \left(  \cdot  \right)$ is a Dirac delta function; $a$ and $b$ are lattice periods along two mutually orthogonal directions that we take to be the $x$- and $y$-axes of the Cartesian coordinate system. On substituting from Eqs. (2) and (3) into Eq. (1) and performing trivial integrations, we obtain
\begin{equation}
\begin{aligned}
&{W^{\left( {out} \right)}}\left( {\bf{r}_1^{'},\bf{r}_2^{'}} \right) = \\
&\sum\limits_{{m_1} = 1}^M {\sum\limits_{{m_2} = 1}^M {{W^{\left( {in} \right)}}\left( {x_1^{'} - {m_1}a,{\rm{ }}y_1^{'} - {m_1}b,{\rm{ }}x_2^{'} - {m_2}a,{\rm{ }}y_2^{'} - {m_2}b} \right)}}.
\end{aligned}
\label{eq4}
\end{equation}

We stress here that each $\delta$-function in Eq. (3) can, in principle, be replaced by any narrow support function such that its widths along the two mutually orthogonal directions are much smaller than those of the CSD ($W^{(in)}$) of the incident beam. The input-output relation of Eq. (4) can be realized in the laboratory to any desired accuracy. We notice that provided the two lattice periods a and b are much larger than the greater of the spot size and coherence width of the incident beam, the cross-terms ($m_1 \neq m_2$) in the sums on the right-hand side of Eq. (4) can be dropped, resulting in the expression
\begin{equation}
{W^{\left( {out} \right)}}\left( {\bf{r}_1^{'},\bf{r}_2^{'}} \right) \approx \sum\limits_{m = 1}^M {{W^{\left( {in} \right)}}\left( {x_1^{'} - ma,{\rm{ }}y_1^{'} - mb,{\rm{ }}x_2^{'} - ma,{\rm{ }}y_2^{'} - mb} \right).} 
\label{eq5}
\end{equation} 
Next, it follows from the DOC definition~\cite{26,27,28} that 
\begin{equation}
{\mu ^{\left( {out} \right)}}\left( {\bf{r}_1^{'},\bf{r}_2^{'}} \right) = \frac{{{W^{\left( {out} \right)}}\left( {\bf{r}_1^{'},\bf{r}_2^{'}} \right)}}{{\sqrt {{I^{\left( {out} \right)}}\left( {\bf{r}_1^{'}} \right)} \sqrt {{I^{\left( {out} \right)}}\left( {\bf{r}_2^{'}} \right)} }}, 
\label{eq6}
\end{equation}
where ${I^{\left( {out} \right)}}\left( \bf{r}^{'} \right) = {W^{\left( {out} \right)}}\left( {\bf{r}^{'}, \bf{r}^{'}} \right)$ is an average output intensity. Notice that the output CSD is a superposition of M non-overlapping terms corresponding to field correlations within M individual nodes of the OCL. It follows that whenever $\bf{r}_1^{'}$ and $\bf{r}_2^{'}$ are situated within the nth node of the lattice, the substitution from Eq. (5) into Eq. (6) yields 
\begin{equation}
\begin{aligned}
{\mu^{\left( {out} \right)}}\left( {\bf{r}_1^{'},\bf{r}_2^{'}} \right) &\approx \frac{{{W^{\left( {in} \right)}}\left( {x_1^{'} - na,{\rm{ }}y_1^{'} - nb,{\rm{ }}x_2^{'} - na,{\rm{ }}y_2^{'} - nb} \right)}}{{\sqrt {{I^{\left( {in} \right)}}\left( {x_1^{'} - na,{\rm{ }}y_1^{'} - nb} \right)} \sqrt {{I^{\left( {in} \right)}}\left( {x_2^{'} - na,{\rm{ }}y_2^{'} - nb} \right)}}} \\
& ={\mu ^{\left( {in} \right)}}\left( {x_1^{'} - na,{\rm{ }}y_1^{'} - nb,{\rm{ }}x_2^{'} - na,{\rm{ }}y_2^{'} - nb} \right).
\end{aligned}
\label{eq7}
\end{equation}
On the other hand, if $\bf{r}_1^{'}$ and $\bf{r}_2^{'}$ are located in different individual nodes of the lattice, the output field is completely uncorrelated at these pairs of points such that 
\begin{equation}
{\mu ^{\left( {out} \right)}}\left( {\bf{r}_1^{'},\bf{r}_2^{'}} \right) = 0.
\label{eq8}
\end{equation}
It then follows that, in general, 
\begin{equation}
{\mu ^{\left( {out} \right)}}\left( {\bf{r}_1^{'},\bf{r}_2^{'}} \right) \approx \sum\limits_{m = 1}^M {{\mu ^{\left( {in} \right)}}\left( {x_1^{'} - ma,{\rm{ }}y_1^{'} - mb,{\rm{ }}x_2^{'} - ma,{\rm{ }}y_2^{'} - mb} \right)} .
\label{eq8}
\end{equation}

The DOC in Eq.(9) has a functional form of a perfect OCL: The spatial node profile of the lattice is determined by the DOC of the incident beam, while the function $P$ controls the overall lattice structure. Eq. (9) is then the main result of this Letter. 

\begin{figure*}
	\includegraphics[width=12.8cm]{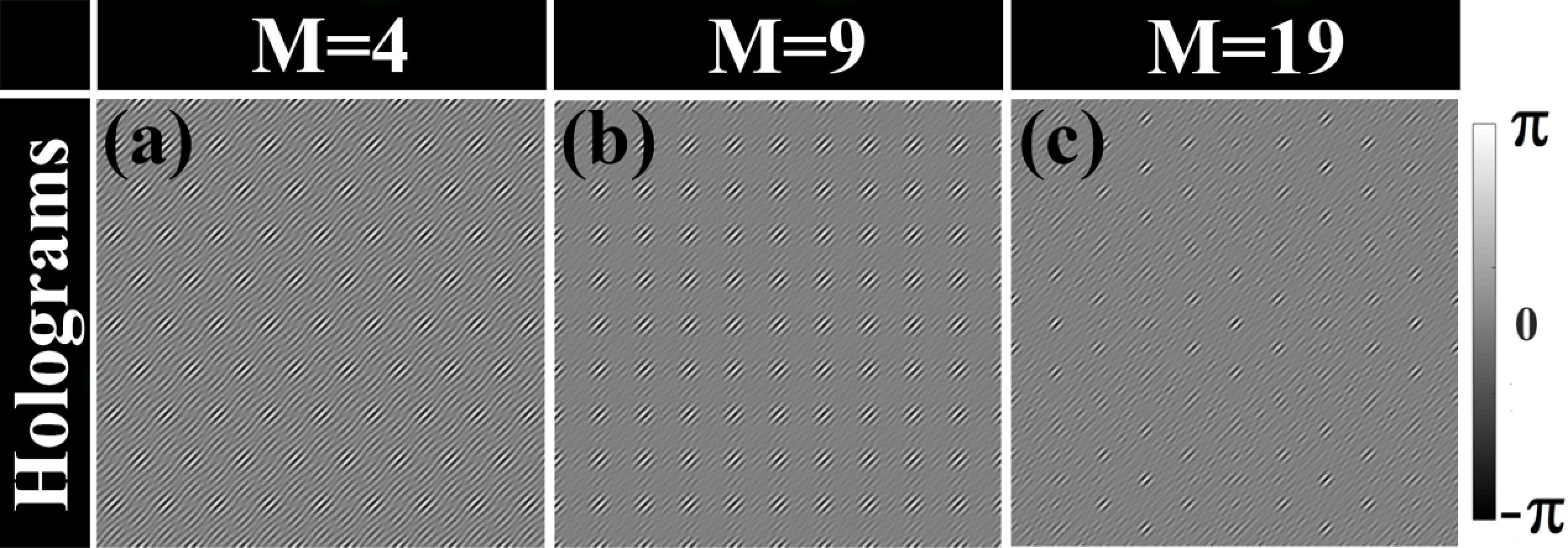}
	\caption{Computer generated holograms of the complex transmission functions of the imaging system producing prefect OCLs.}
	\label{figure4}
\end{figure*} 

\begin{figure*}
	\includegraphics[width=14.8cm]{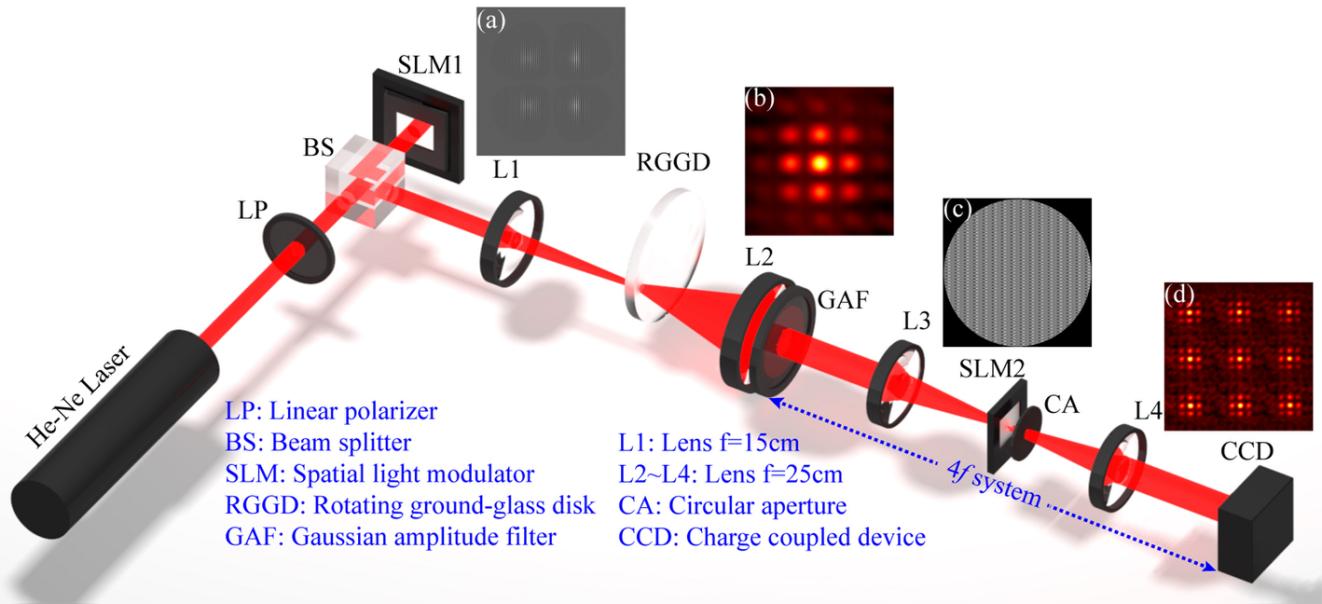}
	\caption{Experimental setup to generate perfect OCLs. The inserts (a)-(d) illustrate the implementation of an OCL with a Hermite-Gaussian node profile. (a): hologram on the SLM1, (b): DOC distribution of the input beam just past a GAF, (c): hologram on the SLM2, (d): OCL profile. }
	\label{figure5}
\end{figure*} 

To demonstrate the feasibility of our method, we first perform numerical simulations for a random input beam of Gaussian intensity and degree of coherence~\cite{26}. We choose an input beam width $\sigma_0$ and its coherence length $\delta_0$ such that $\sigma_0=\delta_0=0.5mm$, and the other parameters are: $f=400mm$ and $\lambda=632nm$. Next, we introduce three complex filters to generate three kinds of perfect OCLs. We exhibit the moduli and phase distributions of the corresponding $P$ functions in Fig.2 (a1)-(c1) and Fig.2 (a2)-(c2), respectively. In Fig. 2 (a3)-(c3), we show the OCLs numerically generated with the aid of these complex transmission functions. The adjacent nodes of each OCL are separated by the distance of 3 mm. Our simulations indicate that the lattice structure can be easily adjusted by varying the lattice periods. Further, to illustrate the OCL node profile control, we studied three Gaussian input beams with the DOC structure generated by Hermite-Gaussian correlated Schell-model (HGCSM) Laguerre-Gaussian correlated Schell-model (LGCSM), and cosine-Gaussian correlated Schell-model (CGCSM) sources. The detailed CSD structure of these sources can be found in Refs.~\cite{28,29,30,31,32}. As a particular example, we display in Fig. 3 the OCLs numerically generated with these three input beams with the help of the complex transmission function with the amplitude and phase shown in Fig. 2(b1) and 2(b2), respectively. We can infer from Figs. 2 and 3 that the OCL structure and node profile can be readily controlled by manipulating the input DOC and the complex transmission function of our imaging system.

We now show how perfect OCLs can be realized in the laboratory. Our protocol hinges on the ability to simultaneously modulate the amplitude and phase of the complex transmission function $P$ in the Fourier plane of our imaging system. Following Refs~\cite{33,34} we can encode the information about the complex transmission function into a phase-only spatial light modulator (SLM). To this end, we inserted the phase only SLM into the Fourier plane and loaded computer generated phase holograms onto it to implement the encoding. In Fig. 4, we exhibit three computer generated holograms corresponding to the complex $P$ functions shown in Fig. 2.

\begin{figure*}
	\includegraphics[width=12.8cm]{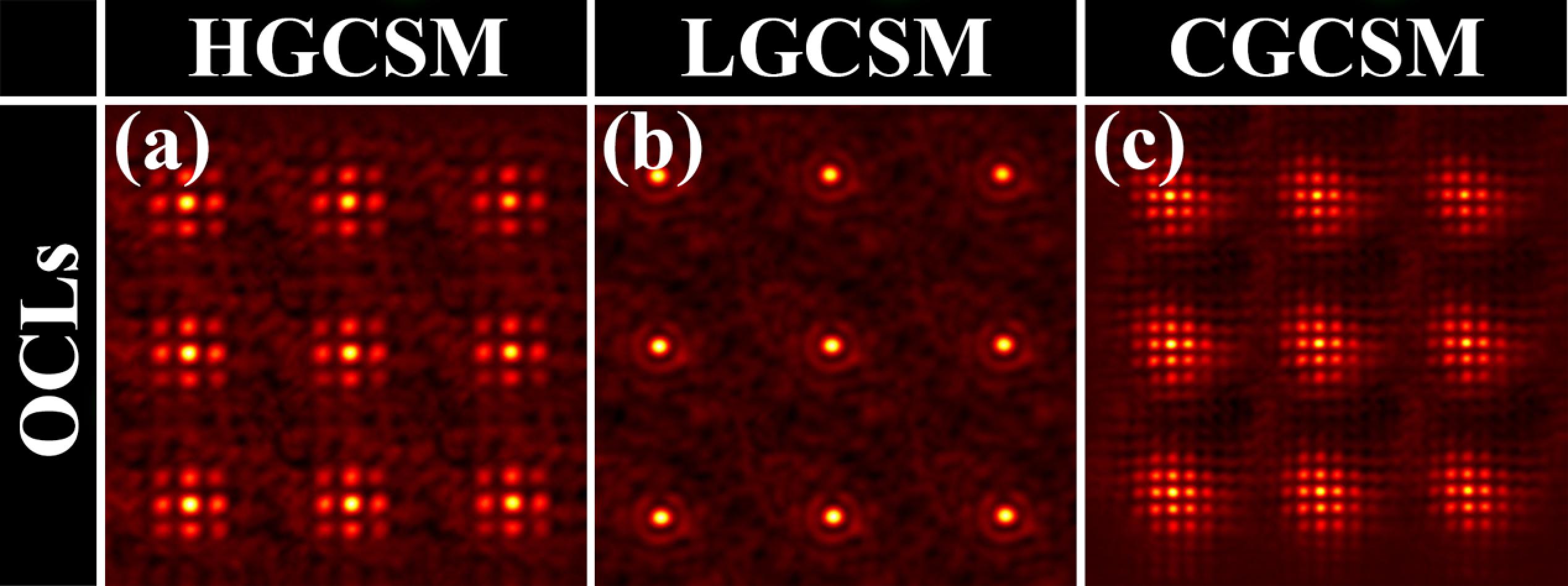}
	\caption{Experimental results for ${\left| {{\mu ^{\left( {out} \right)}}\left( {{\bold{r}_1^{'}},{\bold{r}_2^{'}} = 0} \right)} \right|^2}$ of perfect OCLs. The coherence width and spot size of input partially coherent beams are $\sigma_0=0.35mm$ and $\delta_0=0.17mm$, respectively. The distance between the adjacent nodes equals to 0.73mm.}
	\label{figure6}
\end{figure*} 

We sketch our experimental setup in Fig.5. A He-Ne Laser ($\lambda=632.8nm$) emits a light beam that arrives at the first spatial light modulator (SLM1) after having been transmitted through a linear polarizer (LP) and beam splitter (BS). The SLM1 acts as a phase programmable hologram. The beam emerging from the SLM1 and reflected by the BS, passes through a thin lens L1 and reaches a rotating ground-glass disk (RGGD). We can employ any hologram to engineer a beam intensity profile on the RGGD. The distance between L1 and RGGD is used to control the coherence width of the beam~\cite{31}. Upon transmission through the RGGD, L2 and GAF, a partially coherent input beam with prescribed DOC emerges. The DOC of the emerging beam is given by a Fourier transform of the intensity profile of the beam incident on the RGGD~\cite{31}.  Next, the generated partially coherent beam is focused by a thin lens L3 and illuminates a predesigned hologram on an SLM2. A positive or negative first-order diffraction pattern is then selected by a circular aperture (CA) and transmitted through a thin lens L4. In the rear focal plane of L4, which serves as the output plane to our imaging system, we place a CCD camera to record the random intensity distribution of an output beam. As the light transmitted through the RGGD obeys Gaussian statistics, the Gaussian moment theorem~\cite{27} implies that the DOC of the output beam can be expressed in terms of the intensity correlations as
\begin{equation}
{\left| {{\mu ^{\left( {out} \right)}}\left( {{\bold{r}_1^{'}},{\bold{r}_2^{'}}} \right)} \right|^2} = \frac{{\sum\limits_{n = 1}^N {{I_n}\left( {{\bold{r}_1^{'}}} \right){I_n}\left( {{\bold{r}_2^{'}}} \right)} }}{{NI\left( {{\bold{r}_1^{'}}} \right)I\left( {{\bold{r}_2^{'}}} \right)}} - 1
\label{eq10}
\end{equation}
Here $N$ is a number of ensemble realizations, ${I_n}\left( \bold{r}^{'}  \right)$ is an intensity of the $n^{th}$ realization, and $I\left( \bold{r}^{'}  \right) = \sum\limits_{n = 1}^N {{{{I_n}\left( \bold{r}^{'}  \right)} \mathord{\left/ {\vphantom {{{I_n}\left( \bold{r}^{'}  \right)} N}} \right. \kern-\nulldelimiterspace} N}} $ stands for an average intensity over the ensemble,  see, c.f., Ref.~\cite{32} for more details. In our experiment, the beam spot size and coherence width were measured to be $\sigma_0=0.35mm$ and $\delta_0=0.17mm$, respectively, and the number of ensemble realizations was taken to be $5 \times 10^3$. On comparing Figs. 3 and 6, we can report good agreement between the experiment and numerical simulations, testifying to the practicality of the proposed protocol of perfect OCL generation. 

 In summary, we have introduced the concept of perfect optical coherence lattices and advanced a protocol for their experimental realization.  We have verified our protocol with numerical simulations and implemented it experimentally. Our results open new possibilities for generating customizable perfect optical lattice that are anticipated to find applications to beam splitting, multi-target probing, and free-space optical communications.\\

This work was supported by the National Key Research and Development Program of National Key Research and Development Program of China (2019YFA0705000); National Natural Science Foundation of China (11525418, 11874046, 11947239, 11974218, 91750201, 12004220, 12004225); Natural Sciences and Engineering Research Council of Canada (RGPIN-2018-05497); Innovation group of Jinan (2018GXRC010); China Postdoctoral Science Foundation (2019M662424, 2020M672112).\\

\noindent\textbf{DATA AVAILABILITY}\\
The data that support the findings of this study are available from the corresponding author upon reasonable request. \\

\end{document}